\pdfoutput=1
\documentclass[]{interact}
\usepackage{comment,graphicx}
\usepackage{epstopdf}
\usepackage[caption=false]{subfig}

\usepackage[numbers,sort&compress]{natbib}
\bibpunct[, ]{[}{]}{,}{n}{,}{,}

\theoremstyle{plain}

\theoremstyle{definition}

\theoremstyle{remark}

\newcommand\apj{\emph{The Astrophysical Journal}}%

\newcommand\mnras{\emph{Monthly Notices of the Royal Astronomical Society}}%

\newcommand\aap{\emph{Astronomy and Astrophysics}}

\usepackage{amsmath}
\usepackage{graphicx,psfrag,epsf}
\usepackage{enumerate}
\usepackage{natbib}
\usepackage{url} 
\usepackage{amssymb,amsmath}
\usepackage{color,xcolor,nicefrac}

\def\black{\color{black}}
\begin{document}
\def\DeltaC{$\Delta {C}$}
\def\cstat{${C}$ statistic}
\def\cmin{${C}_{\mathrm{min}}$}
\def\c{$C$}
\def\chandra{\emph{Chandra}}
\def\python{\tt{python}\rm}
\def\pg{PG~1116+215}

\articletype{Original Research Article}

\title{Distribution of the \emph{C} statistic with applications
to the sample mean of Poisson data}

\author{
\name{Massimiliano Bonamente\textsuperscript{a}\thanks{
The author gratefully acknowledges support of \textit{NASA} 
\chandra\ grant AR6-17018X.}}
\affil{\textsuperscript{a} Department of Physics and Astronomy,
University of Alabama in Huntsville, Huntsville AL 35899 (U.S.A)}}
\maketitle

\begin{abstract}
The \cstat, also known as the \emph{Cash} statistic,  is often used in 
astronomy for the analysis 
of low--count Poisson data.
The main advantage of this statistic,
compared to the more commonly used $\chi^2$ statistic,
is its applicability
without the need to combine data points. 
This feature has made the \cstat\ a very useful method to analyze 
Poisson data that have small (or even null) counts in each resolution element.

One of the challenges of the \cstat\ is that
its probability distribution, under the null hypothesis that the 
data follow a parent model, is not known exactly. 
Such distribution is needed for model testing, namely to determine
the acceptability of models and then to determine
confidence intervals of  model parameters.
This is in contrast with the accurate knowledge, for Gaussian data,  of the $\chi^2$ statistic
for any number of free parameters in the parent model.

This paper presents an effort towards improving our understanding of
the \cstat\ by
studying (a) the distribution of \cstat\
for a fully specified model, (b) the distribution of  \cmin\ 
resulting from a maximum--likelihood fit to a simple one--parameter constant 
model, i.e.,  a model that represents the sample 
mean of $N$ Poisson measurements, and (c) the distribution
of the associated $\Delta C$ statistic that is used for parameter
estimation.
The results confirm the expectation that, in the high--count limit, 
both \cstat\ 
and \cmin\ have the same
mean and variance as a $\chi^2$ statistic with same number of degrees of
freedom. It is also found that, in the low--count regime,
the expectation of the \cstat\ and \cmin\ can be substantially 
lower than for a $\chi^2$ distribution. These result have implications
for hypothesis testing in the low--count Poisson regime that are also
discussed in the paper. 

The paper makes use of recent X--ray observations of the astronomical source \pg\
to illustrate the application of the \cstat\ to Poisson data.
These measurements are also used to 
 identify biases in the use
of the $\chi^2$ statistic for Poisson data, especially in the low--count regime.
 
\end{abstract}






\section{Introduction: Advantages and challeges of the \cstat\ for 
modeling low--count Poisson data}
\label{sec:introduction}

Radiation from astronomical sources
is often detected by instruments that collect an integer number of
photons. This is the case for several X--ray and $\gamma$--ray 
instruments \citep[e.g.,][]{bonamente2016}.
Astronomical observations often feature low photon counts, due to a combination
of distance of the source, efficiency of the detectors, and intrinsic luminosity of the source. 
It is customary to combine  the detected counts according to the wavelength
of photons, 
in a number of independent resolution elements or data points.
By such method, a typical astronomical spectrum
is composed of $N$ independent integer measurements $D_i$ ($i$--th data
point),  assumed to
be drawn from a parent Poisson distribution
with unkown mean $S_i$. This is the data model investigated in this paper.

Modeling an astronomical spectrum with a wavelength--dependent function means
first to determine if the data accurately follow the model and, if the model
has adjustable parameters, also to determine such best--fit parameters.
The $\chi^2$ statistic is one of the most used goodness--of--fit statistics 
in astronomy \cite[e.g.,][]{lampton1976}. 
The advantages of the $\chi^2$ statistic is that it can be applied to 
a model with any number of free parameters, and its distribution is
{independent} of the parent means $S_i$.
For a dataset with $N$ independent datapoints and a model with $p$ free parameters,
the best--fit model  has a goodness of fit $\chi^2_{\mathrm{min}}$ that is
distributed like a $\chi^2$ variable with $f=N-p$ degrees of freedom.
This simple property makes hypothesis testing straightforward \citep[e.g.,][]{bevington2003,bonamente2017book}.

The use of the $\chi^2$ fit statistic requires that each 
data point is Gaussian--distributed, and unfortunately its application to low--count
Poisson data 
is not appropriate \citep[e.g.,][]{kaastra2017, humphrey2009,nousek1989}.
For this reason, W.~Cash~\citep{cash1979} introduced
the \cstat\ as a Poisson--based statistic that, like $\chi^2$, is proportional to the logarithm of the 
likelihood.
What is already known is that the \cstat\ is \emph{approximately}
distributed like a $\chi^2$ variable.
The accuracy of this approximation is examined in Sections~\ref{sec:cstat} and  \ref{sec:application-cstat},
where we show that there are significant differences between the two distributions
for small values of the parent Poisson mean.

The mean and variance of the \cstat\ for a fully specified model were
also  studied recently by \cite{kaastra2017}, to show that for a sufficiently
large value of the parent mean, the expectation of $C$ is approximately 1 for each data point.
Use of those results for a model with free parameters is 
however not appropriate. 
When there are $p$ free parameters in the model, a maximum--likelihood method may be
used to determine the statistic \cmin,
which is also 
asymptotically distributed like $\chi^2$ with $N-p$ degrees of freedom.
To date, there has not been a detailed study of the low--count
behavior and the effect of
free parameters on the \cmin\ statistic. This is studied in Sections~\ref{sec:cmin},~\ref{sec:expectation} and \ref{sec:variance},
to show that also \cmin\ has significant differences, 
in the low--count regime, from the $\chi^2_\mathrm{min}$ distribution.

For the purpose of parameter estimation, the likelihood ratio theorem of S.~S.~Wilks~\cite{wilks1938}
shows that the $\Delta C$ statistic can be used for Poisson data \citep{cash1979} 
in much the same way as the $\Delta \chi^2$ statistic 
can be used for Gaussian data \citep[e.g.,][]{lampton1976}, at least in the
asymptotic high--count limit.
Section~\ref{sec:hypothesis} examines the low--count behavior of the $\Delta C$ statistic.
It is found that critical values of \DeltaC\ are consistent with those of 
a $\Delta \chi^2$ distribution, for the simple constant model analyzed in this paper.
At low values of the Poisson mean, there are unique effects due to the
discrete nature of the Poisson distribution that result in differences
between the two distributions.

This paper is structured as follows. Section~\ref{sec:method}
describes the methods of analysis  used to investigate the 
Poisson--based $C$ and \cmin\ statistics.
Section~\ref{sec:application} presents new
theoretical results 
on the distribution of the statistics.
Section~\ref{sec:pg1116} contains an application 
of the $C$ and \cmin\ statistics to
astronomical data of the quasar \pg, with the purpose
of illustrating the use of these statistics on real data.

\section{Methodology}
\label{sec:method}

\subsection{The method of maximum likelihood and the \cstat}
\label{sec:ML-cstat}
The $N$ Poisson data points $D_i$ are assumed to be 
measurements from  a parent model that describes the properties 
of the source. Models
can be either fully specified with no free parameters, or more
commonly  featuring a number of free parameters. 
The likelihood of the data
with the model is
\begin{equation}
\mathcal{L} = \prod_{i=1}^{N} \frac{e^{-S_i} S_i^{D_i}}{D_i!} 
\end{equation}
where  $D_i$ is an integer number of counts (the $i$--th data point) and
$S_i$ the mean value of
the model for that data point. 
It is convenient to calculate the logarithm of the likelihood, 
\[ \ln \mathcal{L} = \sum_{i=1}^N \left( -S_i +D_i \ln S_i - \ln D_i! \right)\]
and then define the
 \emph{Cash} or $C$ statistic as
\begin{equation}
C = -2 \ln \mathcal{L} -B = 2 \sum_{i=1}^N \left( S_i - D_i + D_i \ln (D_i/S_i) \right) = \sum_{i=1}^N C_i
\label{eq:cstat}
\end{equation}
where 
\[ B = 2 \sum_{i=1}^N \left( D_i -D_i \ln D_i + \ln D_i! \right)\]
 and 
\[ C_i = 2 \left( S_i -D_i+D_i  \ln (D_i/S_i) \right). \] 
 The factor of $2$ in the definition of the $C$ statistic 
is introduced for convenience, so that the \cstat\ 
is asymptotically distributed like a $\chi^2$ distribution. 
Notice that $B$ is only 
a function 
of the constant data points $D_i$ and not of the variable model $S_i$, 
and as such it is a constant term that plays no role in the minimization of the likelihood 
\citep{bonamente2017book}. \black
The \cstat\ was introduced by \cite{cousins1984} in the form of Equation~\ref{eq:cstat},
following the initial definition by W. Cash \cite{cash1979}.
 Defining $d=D_i-S_i$ as the deviation of observed counts from the parent model, 
and ignoring terms of the third order in the Taylor series of $\ln (1 +x)$,
\[
        \ln  (S_i/D_i) = \ln \left(1 - \frac{d}{D_i}\right) \simeq - \frac{d}{D_i} - \frac{1}{2} \left(\frac{d}{D_i}\right)^2,
\]
it can be shown that \black the \c\ statistic is approximately equal to the $\chi^2$ statistic: 
\begin{multline}
	{C} \simeq 2 \sum_{i=1}^N \left(S_i - D_i - D_i\left( -\frac{d}{D_i} - \frac{1}{2} \left(\frac{d}{D_i}\right)^2 \right) \right) = \\
	2 \sum_{i=1}^N \left(S_i - D_i +d + \frac{d^2}{2 D_i} \right) = \sum_{i=1}^N \frac{(D_i - S_i)^2}{D_i} = \sum_{i=1}^N \frac{(D_i - S_i)^2}{S_i} \frac{S_i}{D_i}, 
\label{eq:CChi2}
\end{multline}
where, by definition, $S_i - D_i +d =0 $. Since $D_i \sim \text{Poiss}(S_i)$, an 
estimate of the  deviation $d$
is given by the standard deviation of the Poisson distribution, i. e.,
$|d| \simeq \sqrt{S_i}$, further approximated as $|d| \simeq \sqrt{D_i}$, as also suggested by \cite{cash1979}. 
Using this approximation, each term in Equation~\ref{eq:CChi2}
differs from a $\chi^2$ distribution by a factor 
\[
\frac{S_i}{D_i} = \left(1 - \frac{d}{D_i} \right) \simeq  \left(1\pm\frac{1}{\sqrt{D_i}}\right)
\]
which is significant when $D_i$ is small, compared to the expectation
of the other factor in Equation~\ref{eq:CChi2}, $E[(D_i-S_i)^2/S_i]=1$. Even for $D_i=10$, ignoring the $S_i/D_i$ factor
leads to an error of approximately 30\% for each term in the statistic.

For models with adjustable parameters, the parameters are
determined by requiring that $\mathcal{L}$ is maximum for those values.           
Likelihood maximization corresponds to minimization of $- 2 \ln \mathcal{L}$, and
when the maximum--likelihood model parameters are used in the $S_i$ terms of Equation~\ref{eq:cstat},
the statistic takes the name of \cmin. This minimization has an effect to render the
$S_i$ \emph{dependent} on the data $D_i$, as discussed in Section~\ref{sec:cmin}.

\subsection{The \cstat\ for a fully specified model}
\label{sec:cstat}
When the model is fully specified (i.e., with no free parameters), 
the $S_i$ values are known and independent of the data.
The  
null hypothesis that the data are drawn from this parent model means that
\[D_i \sim \text{Poiss}(\mu).
\]
where $\mu=S_i$ is the parent mean of the model.
Under the assumption that the measurements $D_i$ are independent of one another,
 the mean and variance of $C$ can be calculated separately for each data point 
and then summed according to 
\begin{equation}
\begin{cases}
E[C] = \sum_{i=1}^N  E[C_i] \\ 
Var[C] = \sum_{i=1}^N (E[C_i^2] - E[C_i]^2), 
\end{cases}
\label{eq:moments}
\end{equation}
where
\[E[C_i] = 2 \sum_{k=0}^{\infty} \left( (S_i - k + k \ln (k/S_i)) \cdot \dfrac{e^{-S_i} S_i^{k}}{k!} \right)\]
and 
\[
E[C_i^2] = 4 \sum_{k=0}^{\infty} \left( (S_i - k + k \ln (k/S_i))^2 \cdot \dfrac{e^{- S_i} S_i^{k}}{k!} \right)
\]
are respectively the first and second moment of $C_i$, 
the index $k$ representing
all possible values of the Poisson variable $D_i$\black.~\footnote{Throughout this paper,
terms of the type $k \ln k$ or $D_i \ln D_i$ are evaluated
as 0 for $k=0$ or $D_i=0$, as in, e.g.,  \citep{beaujean2011}. }
The two series do not have a simple analytical solution. 
In Section~\ref{sec:application} are derived convenient
numerical approximations for the mean and variance of the \cstat\ according to
these equations, and
it is shown
that the \cstat\ for $N$ datapoints has the same asymptotic  mean and variance
as a $\chi^2$ distribution with $f=N$ degrees of freedom, but significantly smaller
values for small values of the model $S_i$. Expectation and variance of the \cstat\ according to
Equation~\ref{eq:moments} were also reported in \cite{kaastra2017}.

\subsection{The \cmin\ statistic for  a constant model with one free parameter}
\label{sec:cmin}
When the model has free parameters, determined by a fit to the data
 using a maximum likelihood method, the resulting value of the \cstat\
becomes \cmin.
In this case,
the evaluation of expectations according to 
Equation~\ref{eq:moments} is no longer applicable. In fact, the parent mean $S_i$
becomes a statistic
that is now function of the datapoints $D_i$ -- they are no longer fixed numbers, 
as assumed in Equation~\ref{eq:moments}.

The simplest example of this situation is a constant model in which $S_i=S$.
This is the model investigated in this paper,
as an initial study of the effect of free parameters on the \cstat.
For this model, the maximum likelihood method requires that $S$ is in fact the sample mean
of the $N$ measurements \citep{bonamente2017book},
\begin{equation}
S = \frac{1}{N}  \sum_{i=1}^N D_i.
\label{eq:sampleMean}
\end{equation}
Using this sample mean into Equation~\ref{eq:cstat} leads to the \cmin\ statistic
\begin{equation}
C_{\mathrm{min}} = 2 \sum_{i=1}^N D_i \ln \frac{D_i}{S} =  2 \sum_{i=1}^N D_i \ln D_i + 2 M \ln N -2 M \ln M 
\label{eq:cmin}
\end{equation}
where $M = \sum_{i=1}^N D_i$ is the sum of all detected counts.
The $N$ terms $D_i \ln D_i/S$ 
in the sum of Equation~\ref{eq:cmin}\ are no longer independent of each other.
The distribution of \cmin\ is calculated from
the parent distributions of all terms in Equation~\ref{eq:cmin}, namely $ D_i \sim \text{Poiss}(\mu)$ and
 $M \sim \text{Poiss}(N \mu)$, where $\mu$ is the (unknown) parent mean
of $S$.

Compared to the case of the \cstat\ for a fully specified model, the calculation
of moments for \cmin\ is  complicated by the fact that the $N$ terms in Equation~\ref{eq:cmin} 
are \emph{dependent} on one another, because the sample mean $S$ (or the sum of
the counts, $M$) is a function of the data
points $D_i$.
Using an analogy with the $\chi^2$ distribution, 
it is expected that the number of degrees of freedom of the data,
initially $N$, is in fact {reduced} by the presence of free parameters that are fit to
the data to minimize the $C$ statistic. To date, there has not been a study of the effect of
free parameters on the expectation and variance of the \cmin\ statistic. The results presented
in Section~\ref{sec:expectation}
represents a first step in this direction, by calculating how the 
distribution of \cmin\
is modified by the presence of a free parameter,
compared to the case of a fully specified model.


\section{Distribution of the $C$ and \cmin\ statistics}
\label{sec:application}
This section describes theoretical results on 
the distribution of the $C$ and \cmin\ statistics, and the
use of these statistics for hypothesis testing.
The $\Delta C$ statistic is also introduced to 
determine confidence intervals on the model parameters. 

\subsection{Distribution of the  \cstat\ for a fully specified model}
\label{sec:application-cstat}
\begin{figure}[!t]
\centering
\includegraphics[width=5in]{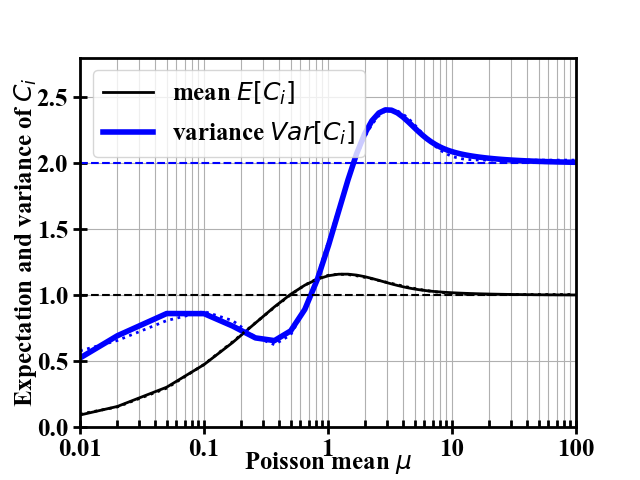}
\includegraphics[width=5in]{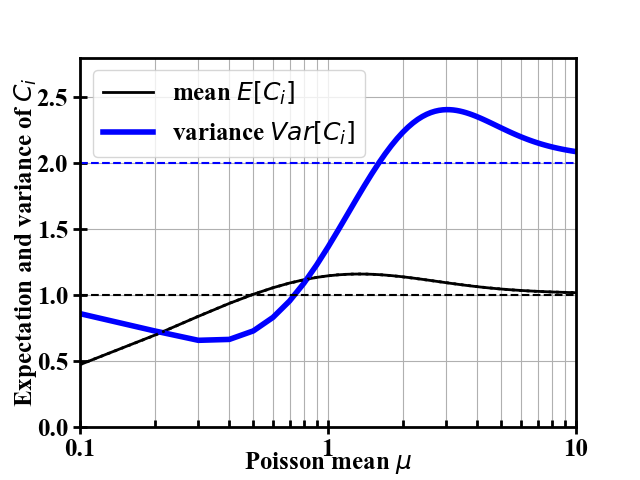}
\caption{(Top) Expectation (in black) and variance (in blue)
of each term of \cstat, according to Equation~\ref{eq:moments}.
Dotted lines are the approximations of Equation~\ref{eq:CFit}
in the range $0.01 \leq \mu \leq 100$.
(Bottom) Same data as in top panel, but with more accurate best--fit models
in the range $0.1 \leq \mu \leq 10$.}
\label{fig:kaastra}
\end{figure}

The first step is the characterization of the \cstat\ for a fully specified model.
 Figure~\ref{fig:kaastra} shows the  mean and variance of each
independent term $C_i$, as a function of
the value of the parent mean $\mu$, obtained from a numerical
evaluation of the series in Equations~\ref{eq:cstat} and \ref{eq:moments}.
Numerical calculations were performed in \emph{python}, using the \emph{scipy}
statistical package. Given the computational challenges associated with the
evaluation of the factorial of large numbers, throughout this paper
the Poisson distribution
is approximated by a Gaussian of same mean and variance for large values
of $\mu$ (see, e.g., Chapter~3 of \cite{bonamente2017book}, for applicability
and accuracy of this approximation). 

Asymptotically,
the numerical calculations reported in Figure~\ref{fig:kaastra} show that 
each term of the $C$ statistic tends to
\[
\lim_{\mu \to \infty} E[C_i] = 1, \;\lim_{\mu \to \infty} Var[C_i] = 2 
\]
and therefore the asymptotic limits for the mean and variance of the $C$ statistic are
\[
\lim_{\mu \to \infty} E[C] = N, \;\lim_{\mu \to \infty} Var[C] = 2N,
\]
which are consistent with the mean and variance of a $\chi^2$ distribution with
a number of degrees of freedom equal to the number of measurements, $f=N$.
This result is expected, since for large values of the Poisson mean,
the Poisson distribution is well approximated by a Gaussian distribution
of same mean and variance, and
the maximum likelihood method applied to $N$ independent Gaussians 
of mean $S_i$ and variance $\sigma_i^2=S_i$ leads to a 
null--hypothesis statistic of
\begin{equation}
\chi^2 = \sum_{i=1}^N \frac{(D_i - S_i)^2}{S_i} ,
\label{eq:chi2}
\end{equation}
i.e., a $\chi^2$ distribution with $N$ degrees of freedom \cite{bonamente2017book}.
Moreover, Equation~\ref{eq:CChi2} shows that the error in approximating ${C}$
with $\chi^2$ is negligible for large Poisson means.

\begin{figure}[!t]
\centering
\includegraphics[width=3in,angle=-90]{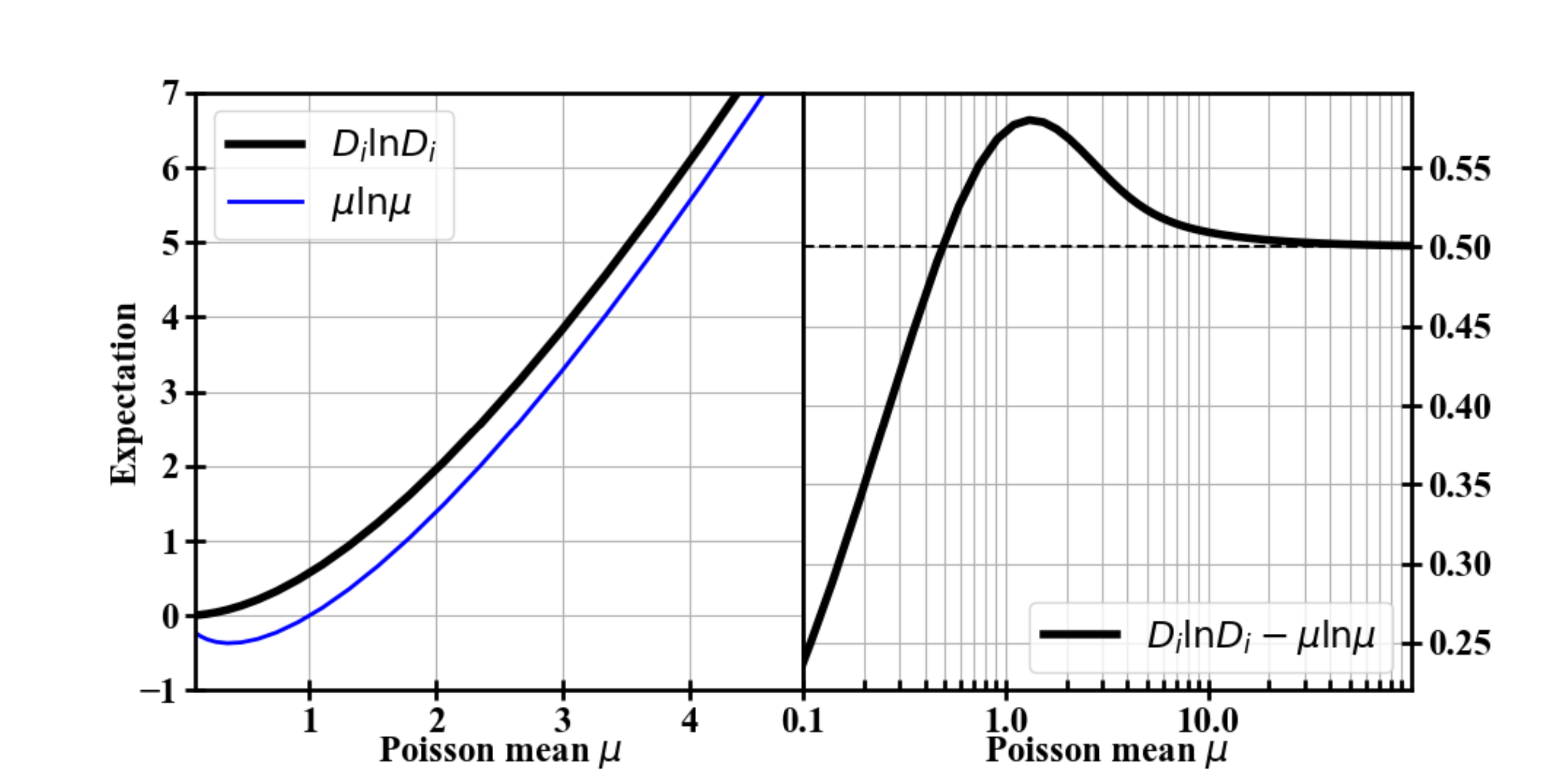}
\caption{(Left): Expectation of function $D_i \ln D_i$ with $D_i \sim P(\mu)$.
Notice that the function $\mu \ln \mu$ is negative
for values of approximately $\mu \leq 1$ before turning positive. (Right):
Expectation of $D_i \ln D_i - \mu \ln \mu$, showing the asymptotic limit
of 1/2 for large $\mu$. }
\label{fig:ENilnNi}
\end{figure}

The asymptotic limit for the expectation of $C_i$ can also
be obtained via a numerical evaluation of 
 the expectation of $D_i \ln D_i$, assuming  $D_i \sim \text{Poiss}(\mu)$.
The expectation can be calculated via
\begin{equation}
E [ D_i \ln D_i] = \sum_{k=0}^{\infty} k \ln k \cdot  e^{-\mu} \frac{\mu^k}{k!},
\label{eq:EDlnD}
\end{equation}
which does not have a simple analytical solution. A numerical solution of this expectation 
as function of the Poisson mean $\mu$ is shown in Figure~\ref{fig:ENilnNi},
with an asymptotic value of
\begin{equation}
\lim_{\mu \to \infty} E[D_i \ln D_i] = \mu \ln \mu+ \frac{1}{2}.
\label{eq:EDlnDlimit}
\end{equation}
With this asymptotic result in hand, the asymptotic limit for the expectation of $C_i$ is
obtained via
\begin{multline}
	\lim_{\mu \to \infty} E[C_i] = \lim_{\mu \to \infty} 2 \cdot E[\mu - D_i + D_i \ln D_i - D_i \ln \mu]= \\
	2 (\mu - \mu + \mu \ln \mu +1/2 - \mu \ln \mu) = 1,
\end{multline}
confirming the result of Figure~\ref{fig:kaastra}.

For convenience, the mean and variance of $C$ can be calculated using the following
approximations:
\begin{equation}
\begin{cases}
E[C_i] = (A + Bx +C(x-D)^2)e^{-\alpha x} +E e^{-\beta x} +F \\
Var[C_i] = (A+Bx^2+C(x-D)^2)e^{-\alpha x} + (E+Fx+G(x-H)^2)e^{-\beta x}+I.
\end{cases}
\label{eq:CFit}
\end{equation}
These approximations are obtained by a fit of the curves of Figure~\ref{fig:kaastra}
with empirical functions that were chosen to describe the two moments accurately and
with just a small number of parameters. 
Values of the parameters $A, B, C, D, E, F, G, H, I, \alpha$ and $\beta$ 
are given in Table~\ref{tab:CFit}; one set of parameters cover the
range $\mu=0.01 - 100$, and another set covers the range $\mu=0.1 - 10$ with better
accuracy (see the two panels in Figure~\ref{fig:kaastra}). 
For small values of the parent means, $\mu \leq 10$, the expectation and variance of \cstat\
differ significantly from the asymptotic values. In particular, the expectation becomes
\emph{significantly smaller than $N$} for approximately $\mu < 1$. This result has
implications for hypothesis testing, as discussed in the following section.

\setlength{\tabcolsep}{4pt}
\begin{table}
\footnotesize
\centering
\caption{Parameters for the functions of Equation~\ref{eq:CFit}.}
\label{tab:CFit}
\begin{tabular}{p{1.1cm}p{1.1cm}p{1.1cm}p{1.1cm}p{1.0cm}p{1.0cm}p{1.1cm}p{1.1cm}p{1.0cm}p{1.1cm}p{1.1cm}}
\hline
A & B & C & D & E & F & G$\times 10^3$ & H & I & $\alpha$ & $\beta$ \\
\hline
\hline
\multicolumn{11}{c}{Parameters for $E[C_i]$ in range $\mu$=0.01--100} \\
0.065672 & -6.9461 & -8.0124 & 0.40165 & 0.261037 & 1.00512 & - & - & - &  
5.5178 & 0.34817  \\
\multicolumn{11}{c}{Parameters for $E[C_i]$ in range $\mu$=0.1--10} \\
-0.56709 & -2.7336 & -2.3603 &  0.52816 & 0.33133 & 1.0174 &- & - & - & 
3.9375 & 0.48446\\
\hline
\multicolumn{11}{c}{Parameters for $Var[C_i]$ in range $\mu$=0.01--100} \\
-2.4637 & 1.5109 & -1.5109 & 0.60509 & 1.4761 & 18.358 & 0.87316
& -0.08592 & 2.02343 & 0.62652 & 7.8187  \\
\multicolumn{11}{c}{Parameters for $Var[C_i]$ in range $\mu$=0.1--10} \\
-3.1971 &  1.5118 & -1.5118 &  0.79384 &  1.9294 &  6.1740 &
  22.360 & -7.2981 & 2.08378 &  0.750315 &  4.49654 \\
\hline
\hline
\end{tabular}
\end{table}

\subsection{Hypothesis testing with the \cstat\ for a fully specified model}
\label{sec:application-cstat-hypothesis}
A discussion of hypothesis testing using the \cstat\ for a fully specified model 
was provided in \cite{kaastra2017}.
In that paper, the author correctly points out that, when there
is a sufficiently large number of independent data points, typically $ N\gg 10$, 
the central limit theorem assures that the \cstat\ has an approximately
 normal distribution. 
According to the central limit theorem, the normal approximation
holds true regardless of the number of counts $D_i$ in each independent data point,
provided that the number of data points $N$ is large. 
The number of counts $D_i$, and their parent mean $\mu$,
has of course an effect on the value of the mean $E[C]$ and variance $Var[C]$, 
as explained in Section~\ref{sec:cstat}. This large--$N$ normal
 approximation is not to be confused with the approximation
of $C$ with a $\chi^2$ variable that occurs in the high--count limit, 
regardless of the value of $N$ (see Section~\ref{sec:ML-cstat}).

Accordingly, in the case of a large number of data points $N$, the normal
approximation to the \cstat\ leads to
a central confidence interval equal to
\begin{equation}
\left[ E[C] - q \sqrt{Var[C]}, E[C] + q\sqrt{Var[C]}\right].
\label{eq:cstatConfidence}
\end{equation}
The parameter $q$ takes values of, e.g., 
 $q=1, 1.7, 2.6$ for, respectively, an enclosed probability $p=0.68, 0.9, 0.99$.
Such central confidence intervals can be used to reject values of \cstat\ that are
too large or too small, according to the null hypothesis.

One--sided confidence intervals are often preferred by data analysists
who choose to reject only large values of the fit statistic.
In that case, one 
defines a critical value of $C$ via $p=\text{Prob}\{C \leq C_{\mathrm{crit}}\}$. When
the \cstat\ can be approximated as normal,
\begin{equation}
C_{\mathrm{crit}} =  E[C] + q \sqrt{Var[C]}
\label{eq:cstatCrit}
\end{equation}
where  values $q=0.5, 1.3, 2.3$ can be used respectively for probabilites $p=0.68, 0.9, 0.99$
\citep[see., e.g., Table A.3 of][]{bonamente2017book}.

The value of the parent Poisson mean $\mu$ of each data point $D_i$ comes into play in 
finding the values of the mean $E[C]$ and variance $Var[C]$.
When the Poisson mean is large, approximately $\mu \geq 10$, 
expressions~\ref{eq:cstatConfidence} and
\ref{eq:cstatCrit}
yield the same results as using a $\chi^2$ distribution, and tables
of critical values of the $\chi^2$ distribution as function of $p$ and $N$
apply to the \cstat\ too \citep[e.g., Table A.3 of ][]{bonamente2017book}. 

On the other hand, when the Poisson mean is small (approximately $\mu \leq 10$),
it is necessary to use the approximations of Equation~\ref{eq:CFit} into \ref{eq:cstatConfidence} and
\ref{eq:cstatCrit}
to calculate confidence intervals and  critical values of the \cstat. 
Such confidence intervals and critical values will differ from those of 
a $\chi^2$ distribution.
In particular, 
the critical value of the \cstat\ can be \emph{substantially smaller than
$N$} when $\mu \ll 1$. For example, a value of $\mu=0.5$ for $N=100$ datapoints leads to a 90\% confidence
value of $C_{\mathrm{crit}} \simeq 60$. For comparison, a $\chi^2$ distribution with the same
number of degrees of freedom has a 90\% critical value of $\chi^2_{\mathrm{crit}} = 118.5$.
Hypothesis testing with a fully-specified \cstat\
must therefore take explicitly into account the value of the parent means of the
Poisson distributions. 

\subsection{Expectation of \cmin\ for the constant model with one free parameter}
\label{sec:expectation}
For models with free parameters, the results of Sections~\ref{sec:application-cstat} and
Section~\ref{sec:application-cstat-hypothesis} are not applicable.
Instead, one must take explicitly into account the fact that
the fit statistic \cmin\ depends on the data through the maximum likelihood
method of minimization, as described in Section~\ref{sec:cmin}.
The relevant statistic becomes \cmin\ according to Equation~\ref{eq:cmin}.

\begin{figure}[!t]
\centering
\includegraphics[width=5in]{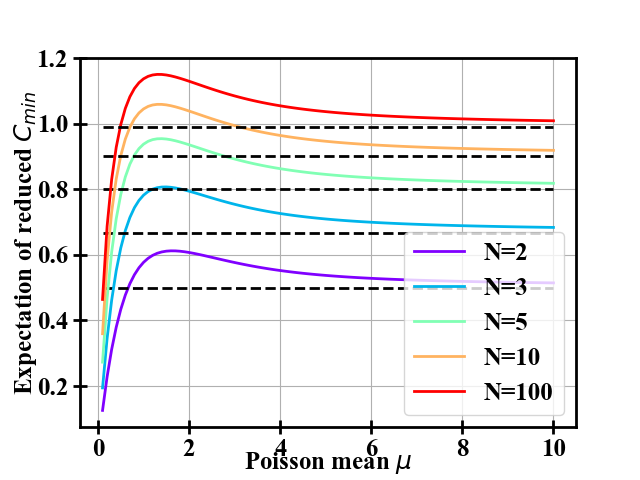}
\includegraphics[width=5in]{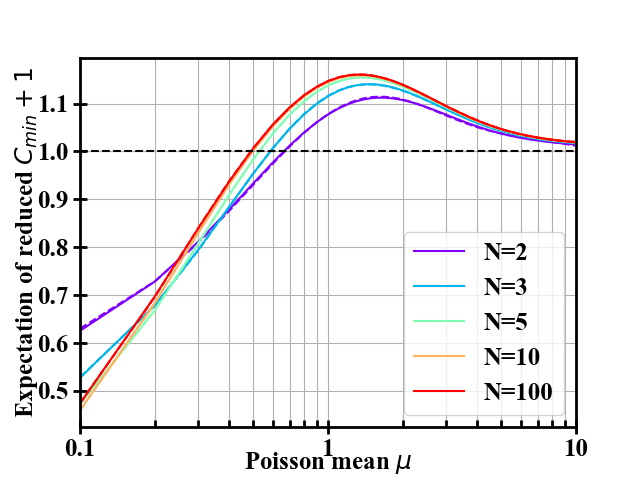}
\caption{(Top) Expectation of \cmin\ 
normalized by the number of variables $N$.
Dashed lines are $(N-1)/N$, corresponding to the asymptotic values of $E[C_{\mathrm{min}}]=N-1$.
(Bottom) Expectation of \cmin+1 divided by $N$. Asymptotic expectation for \cmin\ of
$N-1$ is based on $\chi^2_{\mathrm{min}}$ distribution with $N-1$ degrees of freedom.
For $N > 100$, the curves are virtually indistinguishable 
from the $N=100$ curve. }
\label{fig:ECmin}
\end{figure}

The expectation of \cmin\ according to Equation~\ref{eq:cmin}
can be re--written as
\begin{equation}
E[C_{\mathrm{min}}] = 2 N \cdot E[D_i \ln D_i] + 2 N \mu \ln N - 2 \cdot E[M \ln M].
\end{equation}
where the expectation $E[D_i \ln D_i]$ and its asymptotic value
were presented in Equations~\ref{eq:EDlnD} and \ref{eq:EDlnDlimit}, and 
the expectation $E[M \ln M]$ is carried out in the
same way as $E[D_i \ln D_i]$,  with $M \sim \text{Poiss}(N \mu)$.
The expectation of \cmin\ is reported in Figure~\ref{fig:ECmin}.
In the limit of large $\mu$, the expectation of \cmin\ can be calculated as
\[
\lim_{\mu \to \infty} E[C_{\mathrm{min}}] = 2 N ( \mu \ln \mu + 1/2) + 2 N \mu \ln N 
- 2 ( N \mu \ln N \mu + 1/2 ) = 
N-1,
\]
as shown in Figure~\ref{fig:ECmin}. 
This asymptotic value of $N-1$ for \cmin\ is consistent with the
expectation of the
$\chi_{\mathrm{min}}^2$ distribution. 
In fact, for a one--parameter model, $\chi^2_{\mathrm{min}}$ 
is distributed like a $\chi^2$ distribution with $f=N-1$ degrees of freedom, therefore
with an expectation $E[\chi^2_{\mathrm{min}}] = N-1$  and a variance 
$Var[\chi^2_{\mathrm{min}}] = 2(N-1)$
(see, e.g., \cite{bevington2003} and \cite{bonamente2017book}).
The approximation of Equation~\ref{eq:CChi2} 
also applies to models with free parameters,
showing the asymptotic limit of \cmin\ for one free parameter.

For small values of $\mu$, Figure~\ref{fig:ECmin} shows that 
$E[C_{\mathrm{min}}]$ can be 
significantly different from its 
asymptotic values. This result is qualitatively similar to
the case of the \cstat\ for a fully specified model.
Expectations of the reduced \cmin+1 as function of $\mu$, and for
selected values of $N$, can be accurately represented by the following
empirical function in the range $0.1 \leq \mu \leq 10$,
\begin{equation}
\dfrac{E[C_{\mathrm{min}}]+1}{N} = (A+B\mu +C(\mu-D)^2) e^{-\alpha \mu} + E e^{-\beta \mu} + F
\label{eq:ECminFit}
\end{equation}
The coefficients $A$, $B$, $C$, $D$, $E$,$F$, $\alpha$ 
and $\beta$ \black for Equation~\ref{eq:ECminFit} are reported in 
Table~\ref{tab:ECminFit}.

\begin{table}
\footnotesize
\caption{Model parameters for Equation~\ref{eq:ECminFit}. The $N=100$ parameters
provide a good approximation also for $N > 100$.}
\label{tab:ECminFit}
\begin{tabular}{lcccccccc}
\hline
N & A & B & C & D & E & $\beta$ & F & $\alpha$ \\
\hline
2 & -0.538157 & 0.645002 & 2.230719 $\times 10^{-6}$ & 2.44951 &
 -0.450681 &-8.10319$\times 10^{-3}$ & 1.50052 & 1.21803 \\
3 & -0.992200 & 0.310593 & 1.47508$\times 10^{-4}$&  2.33858&
  0.314182 & 0.490089 & 1.01565 & 2.00042 \\
5 & -0.300655 & -2.29388 & -0.904818 & 0.970195 &  0.331555 &
 0.49173111 &  1.01695262 & 3.45589169 \\
10 & -0.542826 & -2.59328 &-1.99152 & 0.590092 & 
0.332345 & 0.487194& 1.01733 & 3.81537 \\
100 & -0.600716 &-2.66890& -2.360850 & 0.514446&  
0.331258 & 0.484436 & 1.017396 & 3.937691 \\
\hline
\end{tabular}
\end{table}

\subsection{Variance of \cmin\ for the the constant model}
\label{sec:variance}
\begin{figure}
\centering
\includegraphics[width=5in]{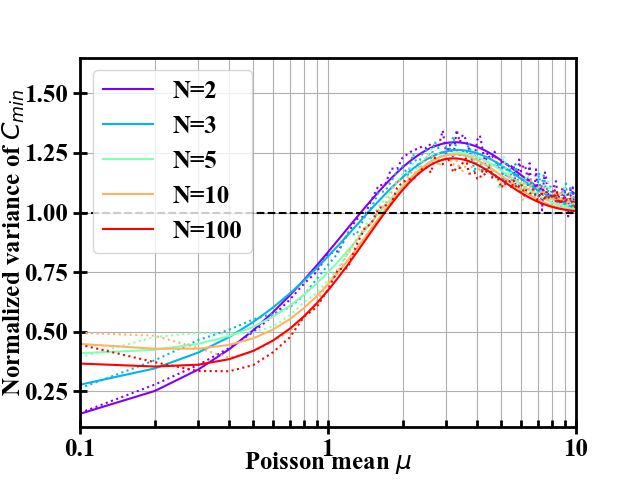}
\includegraphics[width=5in]{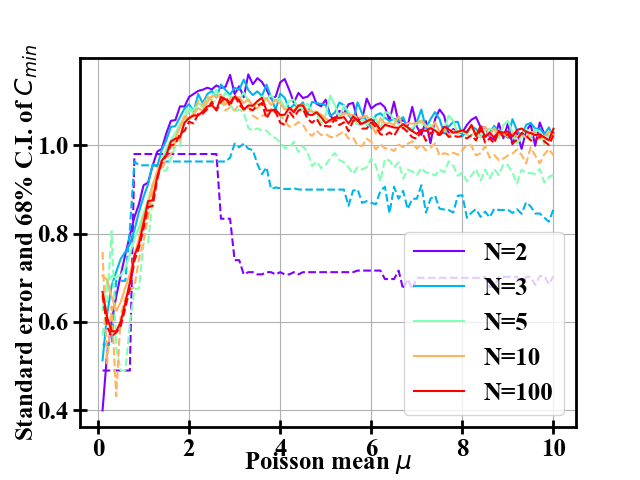}
\caption{(Top) Variance of \cmin\, normalized by $2 (N-1)$.
The  asymptotic expectation of $2(N-1)$ is based on the $\chi^2_{\mathrm{min}}$
distribution for $N-1$ degrees of freedom.
(Bottom) Standard deviation of \cmin\ as solid curves, normalized by the same factor.
Dashed curves are the half--width of central 68\% confidence intervals of
the normalized \cmin. Both sets of curves are normalized by
the same factor $\sqrt{2 (N-1)}$.  
For $N > 100$, the curves are virtually 
indistinguishable from the $N=100$ curve. 
}
\label{fig:variance}
\label{fig:variance68}
\end{figure}

The variance of \cmin\ according to Equation~\ref{eq:cmin} 
cannot be easily calculated analytically.
The main challenge is that $D_i$ and $M$ are correlated,
and therefore the expectations of the products $D_i \ln D_i \cdot M \ln M$ 
do not have a simple analytical form.
The variance is thus estimated via a Monte~Carlo simulation 
of Equation~\ref{eq:cmin}, using 10,000 samples from the $N$
Poisson variables to estimate the resulting variance of \cmin.
Results of the simulations
are show in Figure~\ref{fig:variance}.
As expected, the asymptotic value of the variance is $2(N-1)$,
 equal to
the variance of a $\chi^2$ variable with $N-1$ degrees of freedom,
\[
\lim_{\mu \to \infty} Var(C_{\mathrm{min}}) = 2(N-1).
\]
The variance of \cmin\ is significantly different from the asymptotic value
for small values of the parent mean $\mu$, similar to the
case of the mean.
The variance of \cmin\ can be approximated by the following formula,
\begin{equation}
\dfrac{Var[C_{\mathrm{min}}]}{2(N-1)} = (A+B\mu+C \mu^2) e^{-\alpha \mu} + 1.0,
\label{eq:VarCminFit}
\end{equation}
with the parameters $A$, $B$, $C$ and $\alpha$ provided in Table~\ref{tab:VarCminFit}. 

\begin{table}
\centering
\caption{Model parameters for Equation~\ref{eq:VarCminFit}. The
$N=100$  parameters
provide a good approximation also for $N > 100$.}
\label{tab:VarCminFit}
\begin{tabular}{lcccc}
\hline
N & A & B & C &  $\alpha$ \\
\hline
2 & -0.94444 & 0.38369 & 0.23147 & 0.68654 \\
3 & -0.79062  &-0.12333 & 0.29128   & 0.71632 \\
5 & -0.59153 &-0.54983  & 0.56971 & 0.82521 \\
10 & -0.50551& -1.0592&  0.81869 & 0.90939 \\
100 & -0.59488& -1.0919&  0.85073&  0.94111 \\
\hline
\end{tabular}
\end{table}

The calculations also produce sample distribution functions of \cmin.
For a small number of measurements,
the statistic \cmin\ is necessarily skewed, given the small value of
its mean and  that it is positive definite, similar to the case of
a $\chi^2$ distribution. As a result of this deviation from a normal
distribution, the standard error and the half--width of a central 68\% confidence
interval are not the same, as also shown in Figure~\ref{fig:variance}.
As $N$ increases, the central limit theorem ensures that the distribution
of \cmin\ tends to normal, and  standard error and the half--width 68\% confidence
intervals are in better agreement. The distribution of \cmin\ is needed 
for hypothesis testing, as discussed in the following section.

\subsection{Hypothesis testing and confidence intervals using \cmin}
\label{sec:hypothesis}
The reason to study the distribution of \cmin\ beyond its mean and variance
 is to perform a quantitative
hypothesis testing of the fit, based on $p$--values.
There are two main questions that need to be addressed to test
and use the results of the fit. 
(1)  
Is the value of the fit statistic \cmin\ 
 \emph{acceptable} at a given probability level? 
(2) Assuming that the fit is acceptable, what are 
confidence intervals for the model parameter?

 The term `\emph{acceptable}'
is to be interpreted according to the 
\emph{American Statistical Association}'s statement on $p$--values \citep{asa2016}. 
As is well known, $p$--values simply indicate the degree of \emph{incompatibility} of a data set with
a specified model, and not the probability that the hypothesis specified by
the model is correct.
In other words, $p$--values can only be used to 
\emph{reject} a hypothesis or model
(see also Section~7.1 of \cite{bonamente2017book} 
for further discussion on the subject). 
Accordingly, 
a model is said to be \emph{acceptable} if it \emph{cannot be rejected at a given 
level of probability}, with the understanding that the model may or may not be \emph{the}
correct explanation for the data. An acceptable model is therefore simply 
a plausible explanation that cannot be rejected
by the data at hand, at the probability specified by the $p$--value. 

This section addresses both questions, also discussing the standard
of practice for the use of the more popular $\chi^2$ statistic, and how
such standard of practice can be adapted to \cmin.

\subsubsection{Critical values of \cmin}
\label{sec:hypothesisCmin}
The \emph{acceptability} of a model (in this paper,
a constant model with one parameter) can be addressed by asking whether
the fit statistic \cmin\ is consistent with its parent distribution
based on the null hypothesis that the data are drawn from the
parent model, as studied in Sections~\ref{sec:expectation}
and~\ref{sec:variance}. 

The standard of practice in many fields, including X--ray astronomy
\citep[e.g.,][]{bonamente2016}, is to deem a model \emph{acceptable}
if the fit statistic $\chi^2_{\mathrm{min}}$ (for Gaussian data)
has a value that is less than its critical value, defined via
\[
p = \text{Prob}(\chi^2 \leq \chi^2_{\mathrm{crit}}).
\]
Using the probability ditribution of $\chi^2$,
the critical values are easily evaluated 
(see Table A.7 of \cite{bonamente2017book}). For example,
a one--sided p=90\% confidence interval  with $f=100$ degrees of freedom
has a critical value of $\chi^2_{\mathrm{crit}}=118.5$. Therefore, a fit with
a $\chi^2_{\mathrm{min}} > \chi^2_{\mathrm{crit}}$ should be rejected at the 
90\% confidence level.

\begin{figure}
\centering
\includegraphics[width=5in]{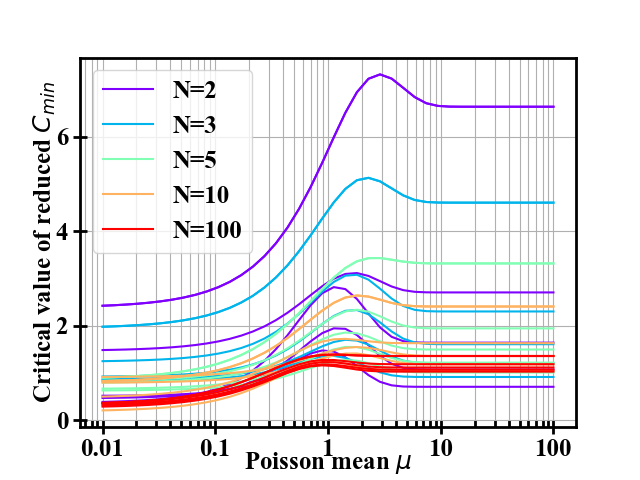}
\caption{Critical values of one--sided confidence intervals for $C_{\mathrm min}/(N-1)$ 
at levels of 60, 70, 80, 90 and 99\% (from bottom to top, for each value of $N$).}
\label{fig:CI}
\end{figure}

This method can therefore be extended to the \cmin\ statistic,
by calculating critical values of \cmin\ at given confidence levels,
defined as
\[
p = \text{Prob}(C_{\mathrm{min}} \leq C_{\mathrm{min, crit}})
\]
Given that the probability
distribution of \cmin\ is not known exactly, the calculation of critical 
values must be carried out by means of the same Monte Carlo simulation
used to estimate the variance of \cmin. In Figure~\ref{fig:CI} are shown
one--sided confidence intervals for \cmin. 
For comparison, equivalent confidence intervals for a $\chi^2$ distribution
with the same number of degrees of freedom are reported in Table~\ref{tab:chi2}.
$N$ is large, critical values can be calculated using the same method
as in Equation~\ref{eq:cstatCrit}, simply replacing the \c\ statistic with the \cmin\
statistic, i.e.,
\begin{equation}
C_{\mathrm{min,crit}} =  E[C_{\mathrm{min}}] + q \sqrt{Var[C_{\mathrm{min}}]}
\label{eq:cminCrit}
\end{equation}
where the meaning of $q$ is described in Section~\ref{sec:application-cstat-hypothesis}.

\begin{table}[!t]
\centering
\caption{Reference values of $\chi^2_{\mathrm{crit}}/f$ for selected values of the 
number of degrees of freedom $f$, reproduced from Table~A.7 of 
\cite{bonamente2017book}.
For comparison, the model presented in this paper has $f=N-1$.}
\label{tab:chi2}
\begin{tabular}{lccccc}
\hline
$f$ & \multicolumn{5}{c}{Null hypothesis probability}\\
	& 0.6 & 0.7 & 0.8 & 0.9 & 0.99\\
\hline
1 & 0.71 & 1.07 & 1.64 & 2.71 & 6.63\\
2 & 0.92 & 1.20 & 1.61 & 2.30 & 4.61\\
4 & 1.01 & 1.22 & 1.50 & 1.95 & 3.32\\
9 & 1.05 & 1.18 & 1.36 & 1.63 & 2.41\\
100 & 1.03 & 1.07 & 1.12 & 1.19 & 1.38\\
$\infty$ & 1.01 & 1.02 & 1.04 & 1.06 & 1.10\\
\hline
\end{tabular}
\end{table}

Comparison between the values of Table~\ref{tab:chi2} and
those in Figure~\ref{fig:CI} show that for large values of the mean ($\mu \geq 5$),
the critical value of \cmin\ are very similar to those of $\chi^2_{\mathrm{min}}$,
for all values of $N$. 
In this range of the parent mean, it is appropriate to use the same critical values
as those of a $\chi^2$ distribution with the same number of degrees of freedom.
For smaller values of $\mu$,  {critical values of \cmin\
can be substantially \emph{smaller} than those based on 
a $\chi^2$ distribution with the same number of degrees of freedom. 
In this range, one expectes substantially smaller values of \cmin\
compared to $\chi^2_{\mathrm{min}}$, and critical values based on the latter
distribution are no longer applicable to \cmin.
This is an important caveat to keep in mind when analyzing 
Poisson data with low count rates.
The results of Figure~\ref{fig:CI} only apply to the simple constant model
analyzed in this paper. One--parameter models with different parameterizations,
or multi--parameter models, may have different critical values, to be calculated
using the appropriate formulation of \cmin\ for those models. Such models
are not discussed in this paper. 

%


\subsubsection{Confidence intervals on the model parameter and the \DeltaC\ statistic}
\label{sec:hypothesisDeltaC}
Another key aspect of model fitting is to determine
a confidence interval for the model parameters.
Methods to determine confidence intervals using
goodness--of--fit statistics were developed
by \cite{lampton1976}, \cite{avni1976} and  \cite{cash1976} using
a method based on the maximum likelihood ratio theorem
proposed by S.~S.~Wilks in 1938~\cite{wilks1938}.
For Gaussian datasets,
confidence intervals on model parameters are calculated using 
critical values of the $\Delta \chi^2$ distribution.
Assuming a one--parameter model as in this paper,
the method is based on the observation that
\begin{equation}
\Delta \chi^2 = \chi^2(N) - \chi^2_{\mathrm{min}}(N-1)
\label{eq:DeltaChi2}
\end{equation}
is distributed like a $\chi^2$ variable with one degree of freedom, where $N$ is the number of
datapoints and $\chi^2_{\mathrm{min}}(N-1)$ is the usual minimum fit statistic.
The $\chi^2(N)$ statistic assumes that the
parameter is fixed at its parent (yet unknown) value, and it has
no free parameters. Under the assumption that
the model remains viable when the parameter value is varied
from its minimum--$\chi^2$ value, $\Delta \chi^2$
is distributed as $\chi^2(1)$.
According to Equation~\ref{eq:DeltaChi2},
a 68\% confidence interval is therefore 
the range that yields $\Delta \chi^2 \leq 1$,
and a 90\% confidence interval corresponds to $\Delta \chi^2 \leq 2.7$.

For Poisson data, W. Cash \cite{cash1979} showed that
the Wilks theorem can be used to generate
confidence intervals for interesting parameters from the statistic
\begin{equation}
\Delta C = C - C_{\mathrm{min}},
\label{eq:DeltaCstat}
\end{equation}
which is howerer only \emph{approximately} distributed like a $\chi^2(1)$ 
distribution. To calculate the exact distribution of \DeltaC,
one
uses Equation~\ref{eq:cstat} with a fixed model
$S=\mu$, i.e,
\[
C = 2 \sum_{i=1}^N \mu - D_i +D_i \ln \left(\frac{D_i}{\mu} \right),
\]
as described in Section~\ref{sec:cstat}.
Unlike $\Delta \chi^2$, it is not possible
to determine exactly the distribution function for $\Delta C$ that
applies to all values of $N$ and $\mu$, and therefore
critical values must be estimated via Monte Carlo simulations of
Equation~\ref{eq:DeltaCstat} as a function of $N$ and $\mu$.  


Critical values of $\Delta C$ are reported in Table~\ref{tab:DeltaC}, along
with the critical values for the reference distribution $\chi^2(1)$.
For small values of $\mu$ and $N$, the probability distribution
function of $\Delta C$ and its critical values reflect the discrete nature of the Poisson distribution.
For example, the probabilities to draw values $D_i$ of respectively 
0, 1, 2 and 3 from a Poisson
with mean $\mu=0.1$ are 90.5\%, 9.05\%,  0.45\% and 0.0015\%.
Therefore, for $N=2$ (first entry in Table~\ref{tab:DeltaC}), most values of \cmin\ are given 
by the following combinations of Poisson draws:

\[
\begin{aligned}
D_i= (0,0) 	&\;C_{\mathrm{min},i} = (0,0)   	& \;C_i = (0.2,0.2) & \; \Delta C = 0.4 \\
D_i= (0,1) 	&\;C_{\mathrm{min},i} = (0,1.4) 	& \;C_i = (0.2,2.8) & \; \Delta C = 1.6 \\
D_i= (1,1) 	&\;C_{\mathrm{min},i} =(1.4,1.4) & \;C_i = (2.8,2.8) & \; \Delta C = 2.8 \\
D_i= (0,2) 	&\;C_{\mathrm{min},i} =(0,2.8) 	& \;C_i = (0.2,8.2) & \; \Delta C = 5.6 \\
\end{aligned}
\]

\begin{table}[!ht]
\centering
\caption{Critical values of $\Delta C$ for 
selected values of the mean $\mu$. Simulations were run
for 10,000 iterations for $\mu=1, 10$  and for 100,000 iterations
for $\mu=0.1$}
\label{tab:DeltaC}
\begin{tabular}{lccccc}
\hline
\hline
$N$   & \multicolumn{5}{c}{Probability $p$}\\
      & 0.6  & 0.7 & 0.8  & 0.9  & 0.99 \\
\hline
\multicolumn{5}{c}{$\mu$=0.1} \\
2 & 0.4 &  0.4 & 0.4 & 1.6 & 5.6  \\
3 & 0.6 &  0.6 & 1.0 & 1.0 & 4.2  \\
5 & 1.0 &  1.0 & 1.0 & 1.0 & 5.8  \\
10 & 2.0 &  2.0 & 2.0 & 2.0& 5.1  \\
100& 0.8 &  1.0 & 1.9 & 3.0& 6.8  \\
\hline
\multicolumn{5}{c}{$\mu$=0.3} \\
2 & 1.2 &  1.2 & 1.2 & 2.0 & 4.9 \\
3 & 1.8 &  1.8 & 1.8 & 1.8 & 5.7 \\
5 & 1.2 &  1.2 & 3.0 & 3.0 & 5.0 \\
10 & 0.4 &  1.1 & 1.8 & 2.3 & 6.0 \\
100 & 0.8 &  1.1 & 1.5& 2.5 & 6.5 \\
\hline
\multicolumn{5}{c}{$\mu$=1.0} \\
2 & 0.6 &  0.6 & 1.5 & 4.0 & 5.2  \\
3 & 0.4 &  1.1 & 1.8 & 2.3 & 6.0  \\
5 & 0.9 &  0.9 & 1.5 & 2.6 & 7.0  \\
10 & 0.8 &  1.0 & 1.9 & 3.0 & 6.8  \\
100 & 0.7 &  1.0 & 1.6 & 2.7 & 6.3  \\
\hline
\multicolumn{5}{c}{$\mu$=3.0} \\
2 & 0.8 &  0.8 & 1.8 & 2.2 & 6.4  \\
3 & 0.9 &  1.1 & 1.6 & 2.4 & 7.0  \\
5 & 0.6 &  1.2 & 1.9 & 2.8 & 7.0  \\
10 & 0.8 &  1.1 & 1.8 & 3.0 & 6.7  \\
100 & 0.7 &  1.1 & 1.7 & 2.7 & 6.7  \\
\hline
\multicolumn{5}{c}{$\mu$=10.0} \\
2 & 0.8 &  1.2 & 1.6 & 2.8 & 7.1  \\
3 & 0.8 &  1.1 & 1.8 & 3.0 & 6.5  \\
5 & 0.7 &  1.0 & 1.7 & 2.6 & 6.6  \\
10 & 0.7 &  1.0 & 1.6 & 2.7 & 6.7  \\
100 & 0.7 &  1.0 & 1.6 & 2.7 & 6.5  \\
\hline
      & \multicolumn{5}{c}{Reference $\chi^2(1)$} \\
\hline 
\bf       &  0.71 & 1.07 & 1.64 & 2.71 & 6.63 \rm \\
\hline
\hline
\end{tabular}
\end{table}

As a result, the critical values of $\Delta C$ follow this set of discrete
values. This discretization of critical values, which
was not present for the $\Delta \chi^2$ statistic, should be taken into
account when performing hypothesis testing with the $\Delta C$ statistic.

For larger values of the mean, $\mu\geq 1$, the estimates
of critical values for $\Delta C$ follow closely those
of a $\chi^2$ distribution for 1 degree of freedom.
This result indicates that it is possible to
treat the $\Delta C$ statistic in the same way as the $\Delta \chi^2$ statistic
for the estimate of confidence intervals on the model parameter.
Therefore, for example, a 90\% confidence interval
of the sample mean is obtained by finding the
range of the sample mean that yield $\Delta C=2.7$, for \emph{any}
value of the sample size $N$.

\section{Applications to X--ray  data of the quasar \pg}
\label{sec:pg1116}
\pg\ is a quasar
located at a distance from Earth of
approximately 
2.5~billion light years and it has been observed by 
the \emph{Chandra} X--ray satellite several times
\citep{bonamente2016,bonamente2019b}.
The X--ray detectors used by \emph{Chandra} collect individual photons 
from the source and neighboring areas in the sky, and measure each
photon's wavelength. The photons are distributed in data points
according to their wavelengths, and 
this  distribution is
usually referred to as the source's \emph{spectrum}. 

\begin{figure}[!t]
\centering
\includegraphics[width=5.5in]{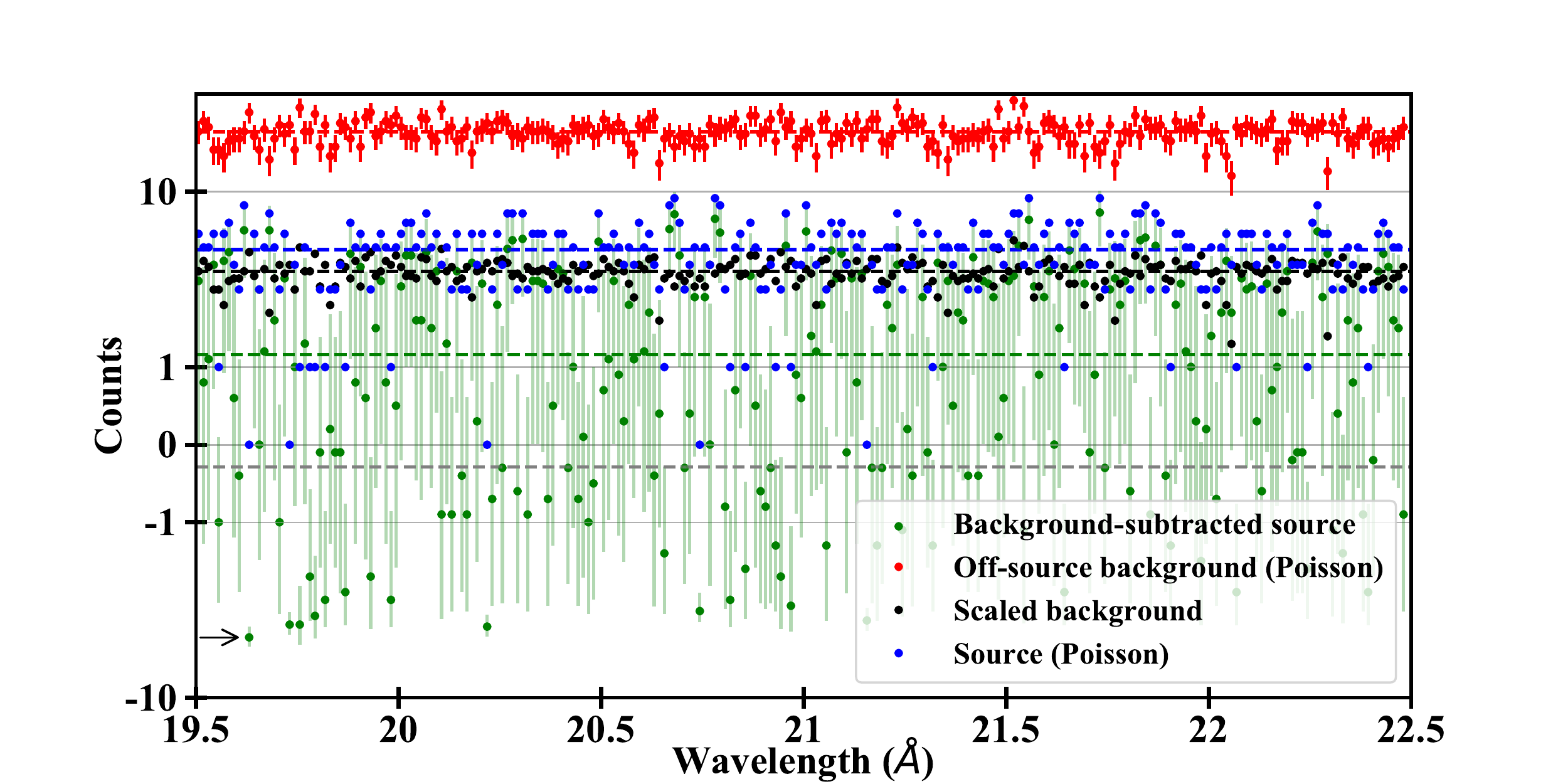}
\caption{ X--ray spectra of \pg\ from an observation taken in 2018 (observation ID 19466).
The off--source background (red, top data set) and the total \pg\ spectrum (blue)
 have  an integer number of counts for each data point.
The scientifically interesting spectrum is the background--subtracted source
spectrum (green).
The point marked by an arrow  and the error bars are discussed in the text. 
(This figure was generated in \texttt{python} using
a \texttt{symlog} scaling for the $y$ axis that enables
a logarithmic representation of numbers around zero.)}
\label{fig:pg1116Spectra}
\end{figure}
One of these X--ray observations 
is used to
illustrate the application of the $C$ and \cmin\ statistics to Poisson data. 
The top curve (red) in Figure~\ref{fig:pg1116Spectra} represent the
spectrum  collected in an off--source region of the detector
that 
is used as a background for  the source spectrum (blue).
These two spectra have $N=239$ independent 
data points with integer counts {that are modelled with a Poisson distribution,} 
in the wavelength range of 19.5--22.5~\AA.~\footnote{The Angstrom (\AA) is 
a unit of measure of length equal to $10^{-10}$~meters.} 
The background spectrum is collected from a larger 
portion of the detector
 than the source area, 
and it is therefore    re--scaled by a deterministic factor $A=0.09992$ 
(black data points) and then subtracted
from the source spectrum to yield the background--subtracted source spectrum (green),
which is the scientifically useful spectrum. 
Since all these spectra are approximately flat in this wavelength range, 
it is appropriate to model them with a constant model, using the methods
described in this paper.~\footnote{The \emph{Chandra} detectors
have a nearly uniform efficiency in this wavelength range. This efficiency is
used to convert the units of `counts' into the scientifically--useful units
of photons per unit time and unit area. Since this correction is nearly uniform
and deterministic,
and this paper does not discuss the astrophysical implications of the spectra, 
the correction is not applied to the spectra.}
\subsection{The off--source background} 
\label{sec:off-source}
The off--source background spectrum 
(Figure~\ref{fig:pg1116Spectra}, red) 
has an average of over 20 counts in the $N=239$ data points.
A maximum--likelihood fit of this Poisson dataset with a constant model yields a best--fit
 mean of $S=26.92$ for a fit statistic of \cmin=232.6, according
to Equations~\ref{eq:sampleMean} and \ref{eq:cmin}. 
According to Figure~\ref{fig:ECmin}, in the high--count regime ($\mu \geq 10$) 
the expectation of \cmin\ is $N-1$, same as for a $\chi^2(N-1)$ distribution.
Critical values of \cmin\ 
can be calculated
from the $\chi^2(N-1)$ distribution, which is followed closely by \cmin\ in this
high--count regime. Using, as an example, a $p$--value of $p=0.9$,
the critical value is approximately
$C_{\mathrm{min,crit}} \simeq 1.1 \times (N-1)=262$, 
as obtained from Table~A.4 of \cite{bonamente2017book}
or from interpolation of the values in Table~\ref{tab:chi2}.
Since the measured value of \cmin\ is lower than the critical value,
the constant model is deemed \emph{acceptable} at the 90\% confidence level.\footnote{
As explained above in Section~\ref{sec:hypothesis}, a model is said to
be \emph{acceptable} if it
cannot be rejected at a specific probability level. It does
 not mean that the model is the correct explanation for the data.}
Other $p$--values can be used following the same procedure.

In X--ray astronomy it is common to fit
data with $\mu \geq 20$ counts, such as this off--source background, using the $\chi^2$ statistic. 
A $\chi^2$ fit 
with a constant model
 yields 
a best--fit mean of $25.91$, with $\chi^2_{\mathrm{min}}=241.34$.
This $\chi^2$ bias towards lower values of the mean, \black 
already noted by others \citep[e.g.][]{humphrey2009},
occurs because the $\chi^2$ best--fit 
is the \emph{weighted} mean of the data points, 
with weights equal to $\sigma_i^{-2}$, instead
of the sample mean 
(for a reference, see Chapter~8 of \cite{bonamente2017book}). 
Since the $\chi^2$ statistic uses $\sigma_i=\sqrt{D_i}$ as 
 an approximation for the standard deviations, 
datapoints with fewer counts 
have larger weight than points with higher counts,
leading to a lower best--fit value.
This bias can be avoided by using the \cmin\ statistic, i.e., retaining the original
Poisson distribution of the data, even in the large--count regime ($\mu \geq 20$).

The distribution of the \c\ statistic (and, for comparison, of
$\chi^2$) around the minimum value of the mean 
is shown in Figure~\ref{fig:pg1116Statistics}.
According to the results of Section~\ref{sec:hypothesisDeltaC}, 
we use a value of $\Delta C = 2.7$ to obtain a 90\% confidence interval for the
mean, and report the best--fit
mean level as $26.92\pm0.54$, at the 90\% confidence level.
If the $\chi^2$ distribution had been used instead, the measurement would have been
(erroneously) biased to a lower value of $25.91\pm0.54$ (using 
$\Delta \chi^2=2.7$ for a 90\% confidence interval).

\begin{figure}
\includegraphics[width=5.5in]{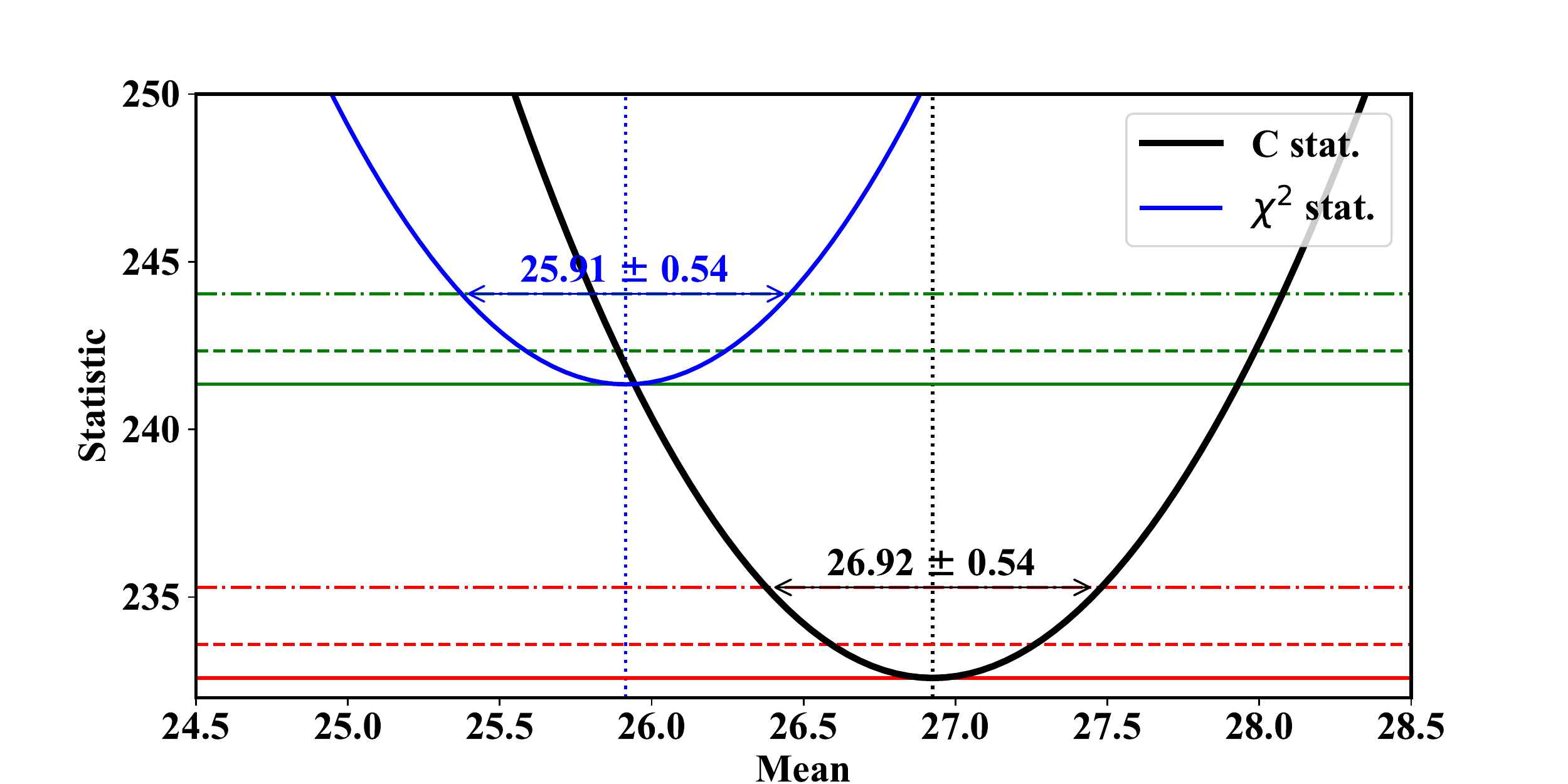}
\caption{ $C$ and $\chi^2$ statistics 
for the off--source background. The minimum values of the statistics 
are marked by solid horizontal lines. Dashed horizontal lines
correspond to \cmin+1 and $\chi^2_{\mathrm{min}}+1$ (for 68\%
confidence intervals on $C$ and $\chi^2$) and 
dot--dashed horizontal lines to \cmin+2.7 and $\chi^2_{\mathrm{min}}+2.7$
(for 90\% confidence).}
\label{fig:pg1116Statistics}
\end{figure}

\subsection{The source spectrum} 
\label{sec:total}
The same method of analysis is applied to the 
source spectrum (Figure~\ref{fig:pg1116Spectra}, blue). 
The fit to a constant model 
returns a best--fit value of
3.85, with a
fit  statistic of \cmin=247.3 for $N=239$ data points.
To test the null hypothesis that the constant model is viable for this spectrum,
the critical value of \cmin\
for $p=0.90$ is needed.
Since this spectrum is in the low--count regime, it is not accurate to
use   Table~A.4 of \cite{bonamente2017book}
or the values in Table~\ref{tab:chi2}, as was done for the off--source spectrum.
Instead, one needs to use Equation~\ref{eq:cminCrit}, which is applicable to
the large--$N$ case, regardless of the value of the Poisson mean.
Equations~\ref{eq:ECminFit}
and \ref{eq:VarCminFit} can be used to estimate the mean and variance of \cmin,
using the coefficients that apply to the $N \geq 100$ case and for a value
of $\mu=3.85$, to obtain $E[C_{\mathrm{min}}]=254.4$ and $Var[C_{\mathrm{min}}]=287.8$.
With these results, the critical value for $p=0.90$ is $C_{\mathrm{min,crit}}=276.5$.
Since $C_{\mathrm{min}} < C_{\mathrm{min,crit}}$, the null hypothesis
cannot be rejected, and the constant model
is  considered acceptable.
A 90\% confidence interval on the best-fit mean is obtained again using
the $\Delta C =2.7$ criterion, and the mean can be reported as $3.85\pm 0.22$.

\subsection{The background--subtracted source spectrum} 
\label{sec:background-subtracted}
The background--subtracted spectrum
is obtained by subtracting a re--scaled version of the off--source background 
(Section~\ref{sec:off-source})
from the source spectrum (Section~\ref{sec:total}). 
To estimate the mean level of emission, 
the non--Poisson nature of this spectrum prevents the direct use of a \cmin\ fit
to this spectrum.
Instead, 
one may combine the results from the analysis of the two Poisson spectra,
which yielded means of respectively $26.92\pm0.54$ (prior to rescaling) and $3.85\pm0.22$.
The background--subtracted mean is therefore $3.85-A \times 26.92 = 1.16$, 
shown as a dashed green line in Figure~\ref{fig:pg1116Spectra}. 
Its uncertainty can be 
estimated by
the error propagation method (see Chapter~4 of \cite{bonamente2017book}), 
and the background--subtracted mean can be reported
as $1.16\pm0.23$ (90\% confidence).

The following illustrates how the low count rate for the source spectrum 
invalidates the use of $\chi^2$ for this background--subtracted source spectrum.
The green error bars represent a Gaussian approximation for the standard deviation $\sigma_i$ of
each background--subtracted data point, calculated according to
\black $\sigma_i^2 = (\sqrt{B_i} A)^2+D_i^2$,
i.e., assuming that both background $B_i$ and source $D_i$ data are Gaussian--distributed.~\footnote{
In fact, the difference of two independent Gaussian variables is a Gaussian
with variance equal to the sum of the two variances, as can be proven as a simple application of the
moment generating function of a Gaussian variable. This method to estimate
the variance is used by the data analysis software that generates the X--ray spectra 
\citep{fruscione2006}.}
This approximation is however \emph{not} accurate because the $D_i$ data
are in the low--count regime where the Poisson distribution is poorly approximated by a Gaussian.
An extreme example is the data point indicated by an arrow near the bottom left of Figure~\ref{fig:pg1116Spectra}
($D_i=0$ and $B_i=37$), where the null contribution to the variance from $D_i=0$ leads to an
estimated background--subtracted rate of $-3.70\pm0.61$, with an artificially small standard deviation.
As a result, using these standard deviations \black
for a $\chi^2$ fit to a constant model results
in a best--fit
mean of -0.28 (the dashed grey line in Figure~\ref{fig:pg1116Spectra})
that is erroneously biased low, compared to the value of $1.16\pm0.23$ obtained earlier. 
The conclusion is that these low--count data should not be fit using the $\chi^2$ statistic,
since the assumption of Gaussian data points is not accurate. 

\subsection{An example of Poisson data with mean $\mu < 1$}
\label{sec:1ks}

A sub--set of the data of Figure~\ref{fig:pg1116Spectra} is used
to further illustrate the application of the \cstat\ 
to low--count data with mean $\mu < 1$.
Figure~\ref{fig:pg1116-1ks} shows the source
 spectrum from a 1,000--second portion of the same observation
shown in Figure~\ref{fig:pg1116Spectra}.
Given the shorter observing time of this sub--set of data,
 the spectrum is mostly composed of 0 or 1 counts per 
data point, with one point having 2 counts. A fit of this data set to
a constant model results in a best--fit mean of $0.100\pm0.034$ 
(90\% confidence 
interval, using $\Delta C = 2.7$), for a fit statistic
of \cmin=113.1.
The small value of the Poisson mean
resulted in a \cmin\ that is substantially smaller than $N-1$.
This point was illustrated
by Figure~\ref{fig:ECmin}, with $E[C_{\mathrm{cmin}}] \simeq \nicefrac{1}{2}\;N$ 
for $N \geq 100$ and $\mu=0.1$,
instead of the asymptotic limit of $E[C_{\mathrm{cmin}}] \simeq N-1$ 
for large values
of the Poisson mean.

The critical value of \cmin\ is calculated as in Section~\ref{sec:total}.
For a value of the Poisson mean of $\mu=0.10$ and $N=239$ data points,
Equations~\ref{eq:ECminFit} and \ref{eq:VarCminFit} yield $E[C_{\mathrm{min}}]=112.4$ 
and $Var[C_{\mathrm{min}}]=85.9$,  and 
Equation~\ref{eq:cminCrit} 
gives a critical value at 90\% confidence of $C_{\mathrm{min,crit}}=124.4$. Since the measured value
of \cmin\ is smaller than the critical value, the constant model is again acceptable
for the data of Figure~\ref{fig:pg1116-1ks}, at the 90\% probability level. 
\begin{figure}[!t]
\centering
\includegraphics[width=5in]{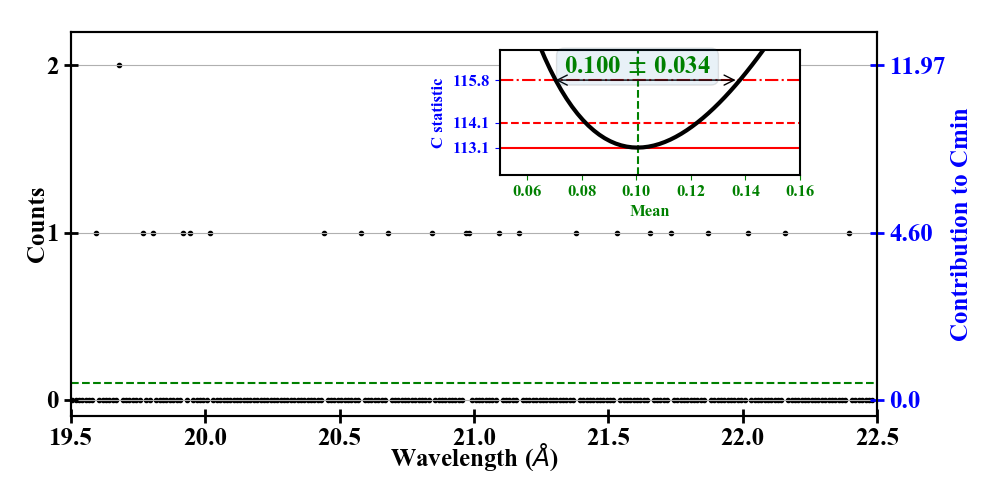}
\caption{Spectrum of \pg\ from a 1,000~second portion of the observation shown in
Figure~\ref{fig:pg1116Spectra}. Green dashed line is the mean value of 0.100 for this
data set, composed mainly of points with zero or one counts. The inset shows the
distribution of \cstat\ around the minimum value.}
\label{fig:pg1116-1ks}
\end{figure}
The vertical axis on the right of Figure~\ref{fig:pg1116-1ks} shows the contribution
of each data point to the \cmin\ statistic. According to
Equation~\ref{eq:cmin}, this contribution is $D_i \ln (D_i/S)$, where 
$S$ is the sample mean. As pointed out in Section~\ref{sec:cstat}, 
data points with $D_i=0$ counts have a null contribution
to the statistic.

\black
\section{Conclusions}

The \cstat\ 
is a goodness--of--fit  statistic derived 
from the maximum likelihood analysis of Poisson data.
It
is the statistic of choice to analyze low--count experiments
such as astronomical spectra from faint sources. 
Challenges in the use of the \cstat\ are associated with the unavailability
of its exact probability distribution function. This paper has provided 
advances in our understanding of the \cstat\ and of the
associated statistics \cmin\ and \DeltaC, their applicability to data analysis,
and identified avenues for future investigations.

First, in the case of a fully specified model, the asymptotic behavior of ${C}$
for a large value of the parent Poisson mean 
is the same as that of a $\chi^2$ distribution with $N$ degrees of freedom, where
$N$ is the number of independent Poisson measurements. 
For smaller values of the Poisson mean, approximately $\mu \leq 10$, there are significant
deviations from the $\chi^2$ distribution. In particular, for $\mu \leq 1$, the expectation
of the \cstat\ is significantly \emph{smaller} than the corresponding $\chi^2$ distribution.
This paper provided convenient analytical approximations for the mean and variance
of the statistic.

Second, this paper investigated the effect of free model parameters by using
a simple constant model with one free parameter, corresponding to the sample mean of the data.
This initial effort showed that the asymptotic behavior of \cmin\
is the same as that of a $\chi^2$ distribution with $N-1$ degrees of freedom.
This result indicates that, just like in the case of the $\chi^2_{\mathrm{min}}$ statistic,
the free parameter in the model has the effect to reduce the number of
degrees of freedom by one. For small values of the Poisson mean, \cmin\
can be substantially different from the corresponding $\chi^2_{\mathrm{min}}$ distribution,
similar to the case of a fully specified model. These results have implications for
hypothesis testing, whereby the statistic has substantially \emph{smaller} critical 
values, compared to a $\chi^2$ distribution with $N-1$ degrees of freedom.

It was also discussed how the  $\Delta C $  statistic can be used to determine confidence intervals
on the model parameter, similar to the case of the $\Delta \chi^2$ statistic. When applied to
this simple constant model, the $\Delta C $ statistic has critical values that are similar to
those of $\Delta \chi^2$, for all $N$. The only caveat in this case is that, for $\mu \leq 1$,
$\Delta C $ has only a few discrete values, unlike $\Delta \chi^2$, and this discretization
may lead to differences between the two statistics. The indication provided by this study is that,
within the uncertainties due to the discrete nature of the \cstat, $\Delta C$ can be used in
much the same way as $\Delta \chi^2$.

The application of these methods to observations of the quasar \pg\ showed that the \cmin\
statistics is to be preferred to the $\chi^2_{\mathrm{min}}$ statistic to fit
counting (i.e., Poisson--distributed) data. In the low--count regime, the 
$\chi^2_{\mathrm{min}}$ statistic is not applicable because the data cannot
be accurately approximated by a Gaussian distribution. Even in the high--count
limit, the $\chi^2_{\mathrm{min}}$ statistic was shown to bias the constant
model to lower best--fit values, compared to the \cmin\ statistic. The findings of this
paper therefore support the recommendation that the Poisson--based $C$ and \cmin\
statistics be the statistics of choice to fit counting data, regardless of the number of counts. 

The results provided in this paper may not be directly applicable to models with
more than one free parameter, or for one--parameter models
other than the simple constant model. An extension of this study to such models
is required to ensure the applicability of these results to more complex models,
in particular with regards to the result that \cmin\ is significantly smaller
than $\chi^2$ in the small--count regime.
For models with $p > 1$ free parameters, it is known that the asymptotic distribution 
of \cmin\ tends
to that of a $\chi^2$ distribution with $N-p$ number of degrees of freedom, consistent with
the results of this paper. For such multi--parametric models, it is necessary to 
investigate further
both the low--count behavior and whether the parameterization of the model has an effect of
the distribution of \cmin.

\bibliographystyle{apj}

\end{document}